\documentclass{article}

\usepackage{arxiv}

\usepackage[utf8]{inputenc} 
\usepackage[T1]{fontenc}    
\usepackage{hyperref}       
\usepackage{url}            
\usepackage{booktabs}       
\usepackage{amsfonts}       
\usepackage{nicefrac}       
\usepackage{microtype}      
\usepackage{lipsum}		
\usepackage{graphicx}
\usepackage{amsmath}
\usepackage{doi}

\title{SEA-ViT: Sea Surface Currents Forecasting Using Vision Transformer and GRU-Based Spatio-Temporal Covariance Modeling}

\author{ \href{https://orcid.org/0000-0001-8464-4476}{\includegraphics[scale=0.06]{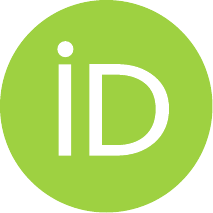}\hspace{1mm}Teerapong Panboonyuen}\thanks{I thank myself for making this work possible, hoping it helps improve vision-based models and inspires others. Explore more about me at \href{https://kaopanboonyuen.github.io/}{https://kaopanboonyuen.github.io/}.} \\
	Postdoctoral Researcher, Chulalongkorn University\\
	Senior Research Scientist, MARS (Motor AI Recognition Solution)\\
	\texttt{teerapong.panboonyuen@gmail.com} \\
}

\renewcommand{\shorttitle}{Teerapong Panboonyuen}

\hypersetup{
pdftitle={SEA-ViT},
pdfsubject={AI},
pdfauthor={Teerapong Panboonyuen},
pdfkeywords={Computer Vision, AI, Deep Learning},
}

\begin{document}
\maketitle

\begin{abstract}  
Forecasting \textbf{sea} surface currents is essential for applications such as maritime navigation, environmental monitoring, and climate analysis, particularly in regions like the Gulf of Thailand and the Andaman Sea. This paper introduces \textbf{SEA-ViT}, an advanced deep learning model that integrates Vision Transformer (ViT) with bidirectional Gated Recurrent Units (GRUs) to capture spatio-temporal covariance for predicting \textbf{sea} surface currents (U, V) using high-frequency radar (HF) data. The name \textbf{SEA-ViT} is derived from \textbf{Sea} Surface Currents Forecasting using \textbf{Vi}sion \textbf{T}ransformer, highlighting the model's emphasis on ocean dynamics and its use of the ViT architecture to enhance forecasting capabilities. \textbf{SEA-ViT} is designed to unravel complex dependencies by leveraging a rich dataset spanning over 30 years, incorporating ENSO indices (El Niño, La Niña, and neutral phases) to address the intricate relationship between geographic coordinates and climatic variations. This development enhances the predictive capabilities for \textbf{sea} surface currents, supporting the efforts of the Geo-Informatics and Space Technology Development Agency (GISTDA) in Thailand's maritime regions. The code and pretrained models are available at \href{https://github.com/kaopanboonyuen/gistda-ai-sea-surface-currents}{https://github.com/kaopanboonyuen/gistda-ai-sea-surface-currents}.
\end{abstract}

\section{Introduction}

Understanding \textbf{sea} surface currents is critical for various maritime applications, such as navigation, fisheries management, and climate modeling. These currents play a pivotal role in shaping marine ecosystems and influencing human activities in both coastal and open ocean environments. Traditional methods for extracting surface current vectors from High-Frequency (HF) radar data often rely on deterministic models, which struggle to capture the complex, non-linear spatio-temporal dependencies inherent in ocean current dynamics \cite{li2023deep}.

In response to these challenges, we introduce \textbf{SEA-ViT} (short for \textbf{S}ea Surface Currents Forecasting Using \textbf{Vi}sion \textbf{T}ransformer and GRU-Based Spatio-Temporal Covariance Modeling), an advanced deep learning framework that integrates the Vision Transformer (ViT) \cite{dosovitskiy2020image} with bidirectional Gated Recurrent Units (GRUs) to model the spatio-temporal covariance of sea surface currents. The name \textbf{SEA-ViT} reflects both the model's focus on \textbf{sea} surface currents and its use of the \textbf{Vi}sion \textbf{T}ransformer architecture to enhance forecasting accuracy.

The ViT, originally designed for image-based tasks, has demonstrated strong capabilities in capturing global dependencies and structural patterns in data. By applying this architecture to sea surface current forecasting, \textbf{SEA-ViT} can effectively model long-range spatial interactions, which are crucial for accurately predicting ocean dynamics \cite{dosovitskiy2020image, chang2021real}. The ViT's self-attention mechanism allows the model to focus on the most relevant spatio-temporal features across the sea surface, providing a significant advantage over traditional convolutional models that have limited receptive fields.

This novel approach is specifically designed to predict the U and V vector components of sea surface currents with high accuracy. The significance of this model is particularly notable for Thai waters, including the Gulf of Thailand and the Andaman Sea, where precise current forecasting is essential for effective maritime operations and environmental management. By leveraging the global attention capabilities of the Vision Transformer alongside the temporal memory strengths of GRUs, \textbf{SEA-ViT} overcomes the limitations of traditional models and provides a robust tool for understanding and predicting sea surface currents in these critical regions.

Additionally, the inclusion of the ENSO index, which accounts for oceanic changes driven by climate phenomena like El Niño and La Niña, enhances the model's predictive ability by capturing both short- and long-term dependencies across space and time. This integration makes \textbf{SEA-ViT} well-suited to handle the dynamic and complex nature of sea surface currents, supporting a wide range of maritime and environmental applications.

\section{Data Handling and Preprocessing}

The dataset for predicting sea surface currents includes historical HF radar measurements \cite{smith2021time}, which are represented as vectors \( (U, V) \), geographical coordinates (latitude, longitude), and timestamps (datetime). Additionally, the ENSO index is used to account for climate-induced variations. The ENSO index is categorical, with values indicating neutral (0), El Niño (1), and La Niña (2) conditions.

\subsection{Data Splitting}

The dataset is partitioned into three subsets: training, validation, and testing. This split ensures that the model is evaluated on unseen data and helps in tuning the hyperparameters effectively. The proportions of the split are typically 70\% training, 15\% validation, and 15\% testing, though these may vary based on the dataset size and specific experimental needs.

\subsection{Normalization of Sea Surface Current Vectors}

Normalization is a crucial preprocessing step that adjusts the data to a common scale, improving the convergence and performance of the machine learning model. For sea surface current vectors \( U \) and \( V \), we use standard score normalization (z-score normalization), defined by:

\begin{align}
    u' &= \frac{u - \mu_u}{\sigma_u}, \\
    v' &= \frac{v - \mu_v}{\sigma_v}
\end{align}

where:
- \( u \) and \( v \) are the original sea surface current components in the eastward and northward directions, respectively,
- \( \mu_u \) and \( \mu_v \) are the mean values of \( u \) and \( v \) over the training set,
- \( \sigma_u \) and \( \sigma_v \) are the standard deviations of \( u \) and \( v \) over the training set,
- \( u' \) and \( v' \) are the normalized sea surface current components.

\subsubsection{Mathematical Rationale}

Normalization is applied to ensure that the input features \( U \) and \( V \) have zero mean and unit variance. This scaling is particularly beneficial for models that rely on gradient-based optimization techniques. The z-score normalization transforms the data into a distribution with a mean of 0 and a standard deviation of 1, which standardizes the influence of each feature on the model. The benefit of this transformation in the context of predicting sea surface currents includes:

\begin{itemize}
    \item Improved Convergence:
    \begin{itemize}
        \item By standardizing the input features, the gradients computed during backpropagation are scaled similarly, which often leads to faster and more stable convergence of the optimization algorithm.
    \end{itemize}
    
    
    \item Better Performance:
    \begin{itemize}
        \item Normalization can help in achieving better performance metrics as it ensures that the model treats all features on an equal footing during training.
    \end{itemize}
\end{itemize}

\subsection{Integration with Temporal and Spatial Features}

The normalized sea surface current vectors \( U \) and \( V \) are combined with other features such as latitude, longitude, datetime, and the ENSO index. This integration allows the model to account for the spatial and temporal context of the sea surface currents:

\begin{itemize}
    \item Temporal Features: 
    \begin{itemize}
        \item The datetime information is crucial for capturing temporal dependencies and variations in sea surface currents.
        \item Normalization of \( U \) and \( V \) ensures that these temporal patterns are not distorted by variations in the scale of the current vectors.
    \end{itemize}
    
    \item Spatial Features: 
    \begin{itemize}
        \item Latitude and longitude provide spatial context.
        \item Although these features are not normalized, their integration with the normalized \( U \) and \( V \) vectors allows the model to capture geographical influences on current patterns.
    \end{itemize}
    
    \item Climate Features: 
    \begin{itemize}
        \item The ENSO index represents large-scale climate phenomena and is used as an external input to account for significant climate-induced variations.
        \item It complements the normalized \( U \) and \( V \) vectors by providing context for broader climate impacts.
    \end{itemize}
\end{itemize}

In summary, the normalization of sea surface current vectors \( U \) and \( V \) is a key preprocessing step that standardizes the data, improving the model's efficiency and performance. By applying this technique, we ensure that the model can effectively learn and predict the complex dynamics of sea surface currents in Thailand's waters.

\section{Model Architecture}

The proposed architecture integrates bidirectional Gated Recurrent Units (GRUs) for capturing temporal dependencies with a transformer-based self-attention mechanism for modeling spatial interactions. This hybrid approach leverages the strengths of sequential and spatial feature processing to enhance the prediction accuracy of sea surface current vectors.

\begin{figure}[ht]
  \centering
  \includegraphics[width=0.8\linewidth]{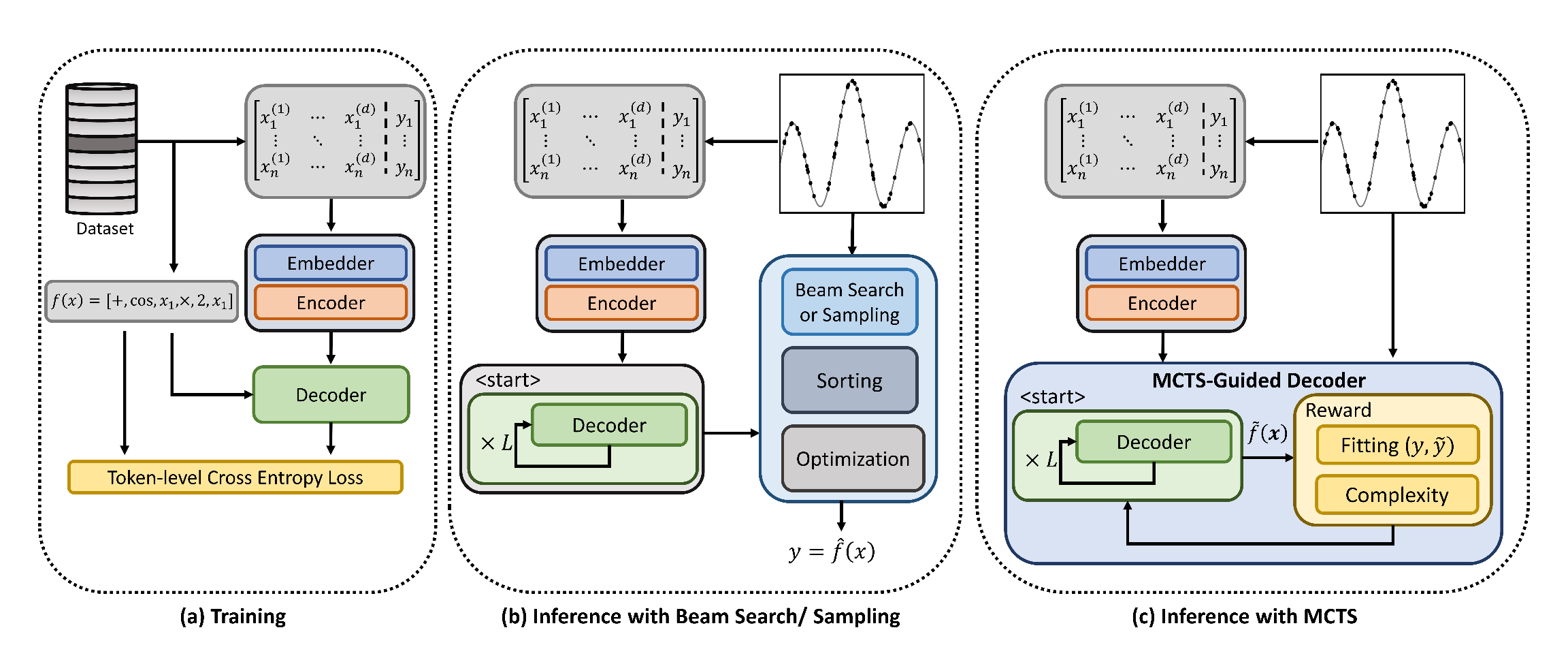}
  \caption{Proposed GRU-Transformer architecture for predicting sea surface current vectors. This framework is inspired by the transformer-based planning model for symbolic regression presented by Shojaee et al. (2024) \cite{shojaee2024transformer}.}
\label{GISTDA_PP_01}
\end{figure}

\subsection{Bidirectional GRU Layer}

Bidirectional GRUs extend the capacity of traditional GRUs by processing input sequences in both forward and backward directions. This bidirectional processing allows the model to capture temporal dependencies that span both past and future contexts, which is crucial for understanding the dynamics of sea surface currents.

The bidirectional GRU update and reset gates are defined as follows:

\begin{align}
    z_t &= \sigma(W_z x_t + U_z h_{t-1} + b_z), \\
    r_t &= \sigma(W_r x_t + U_r h_{t-1} + b_r),
\end{align}

where:
- \( \sigma(\cdot) \) denotes the sigmoid activation function,
- \( W_z \) and \( U_z \) are weight matrices for the update gate,
- \( W_r \) and \( U_r \) are weight matrices for the reset gate,
- \( b_z \) and \( b_r \) are bias terms.

The hidden state \( h_t \) is updated by combining the previous hidden state and the current input through:

\begin{align}
    h_t &= (1 - z_t) \circ h_{t-1} + z_t \circ \tanh(W_h x_t + U_h (r_t \circ h_{t-1}) + b_h),
\end{align}

where:
- \( \circ \) denotes element-wise multiplication,
- \( W_h \) and \( U_h \) are weight matrices for the hidden state update,
- \( b_h \) is the bias term.

Bidirectional GRUs are essential for this model as they enable the learning of temporal patterns in the sea surface currents, such as seasonal and cyclical variations, from both past and future data points.

\subsection{Self-Attention Layer}

The self-attention mechanism in the transformer model allows the network to focus on different parts of the input sequence, capturing dependencies across both time and space. The attention mechanism is defined mathematically by:

\begin{align}
    \text{Attention}(Q, K, V) &= \text{softmax}\left(\frac{QK^T}{\sqrt{d_k}}\right)V
\end{align}

where:
- \( Q \) (query), \( K \) (key), and \( V \) (value) are matrices derived from the input features,
- \( d_k \) is the dimension of the key vectors.

The scaled dot-product attention computes the relevance of each query with all keys, adjusting the influence of each value based on these relevances. Specifically, the attention weights are computed as:

\begin{align}
    \text{Attention}_{i,j} &= \frac{\exp\left(\frac{Q_i K_j^T}{\sqrt{d_k}}\right)}{\sum_{k} \exp\left(\frac{Q_i K_k^T}{\sqrt{d_k}}\right)},
\end{align}

where \( Q_i \) and \( K_j \) are the query and key vectors for the \( i \)-th and \( j \)-th positions, respectively.

This attention mechanism enables the model to dynamically focus on relevant spatial features, such as latitude and longitude, and their interactions with temporal components, including datetime and ENSO index. It allows the model to capture complex spatial dependencies and temporal correlations in sea surface currents.

In this work, we extend the GRU-Transformer architecture to predict the sea surface current vectors \( U \) and \( V \). The bidirectional GRU layer processes the temporal sequences of input features, capturing the evolution of sea surface currents over time. The transformer self-attention layer models the spatial correlations and interactions between features, enhancing the prediction accuracy for \( U \) and \( V \) components.

By combining these components, our model leverages both temporal and spatial information, providing a robust framework for predicting sea surface currents in Thailand waters.

\section{Training Procedure}

The training procedure involves multiple steps, starting from data preprocessing to model deployment via MLOps integration. Below is a detailed breakdown:

\subsection{Data Overview and Input Format}

The input dataset for sea surface currents prediction includes several key features such as temporal (`datetime`), spatial (`lat`, `lon`), and vector components of the sea surface currents (`u`, `v`), along with the ENSO (El Niño Southern Oscillation) index. These features play a crucial role in the model's ability to predict sea surface currents accurately, as they capture both the physical properties and the climate-related patterns that influence the ocean's dynamics. The ENSO index, in particular, is widely acknowledged as a significant climate variable impacting oceanic currents, making it a critical feature in models that account for long-term climatic variations (e.g., El Niño and La Niña events) \cite{kim2023overemphasized}.

A sample of the input data format is provided in Table \ref{tab:input_data}. Each row in the dataset represents a specific timestamp and geographical location, characterized by the latitude and longitude coordinates. The variables \(U\) and \(V\) denote the eastward and northward components of the sea surface current velocity, respectively, measured in meters per second (m/s). The `ensoindex` column reflects the ENSO phase during the observation, with values indicating neutral, El Niño, or La Niña conditions.

\begin{table}[ht!]
\centering
\caption{Sample of input data for sea surface currents prediction.}
\label{tab:input_data}
\begin{tabular}{@{}llllll@{}}
\toprule
\textbf{Datetime}          & \textbf{Lat}   & \textbf{Lon}   & \textbf{U}      & \textbf{V}       & \textbf{ENSO} \\ \midrule
2024-01-01 00:01:00 & 9.0049 & 99.9213 & 0.0563 & -0.2096 & 0 \\
2024-01-01 00:02:00 & 9.0049 & 99.9368 & 0.1401 & -0.6546 & 0 \\
2024-01-01 00:03:00 & 9.0049 & 99.9467 & 0.1510 & -0.6725 & 1 \\
2024-01-01 00:04:00 & 9.0049 & 99.9573 & 0.0853 & -0.5897 & 1 \\
2024-01-01 00:05:00 & 9.0049 & 99.9153 & 0.1621 & -0.7715 & 2 \\
2024-01-01 00:06:00 & 9.0049 & 99.9243 & 0.0918 & -0.6139 & 2 \\
\bottomrule
\end{tabular}
\end{table}

\subsection{Detailed Analysis of Input Data}

The dataset captures the dynamic behavior of sea surface currents over time and space. The temporal granularity, provided by the `datetime` feature, ensures that changes in current velocities due to diurnal or seasonal variations can be modeled. Moreover, the geographical coordinates (`lat`, `lon`) allow the model to learn spatial dependencies, which is crucial for capturing localized oceanic phenomena such as gyres, upwelling, and coastal boundary currents.

\subsubsection{Datetime}

The **Datetime** feature represents the specific time at which the sea surface current measurement is recorded. Since ocean currents are influenced by both short-term fluctuations (such as tidal cycles) and long-term changes (such as seasonal shifts and climate anomalies), capturing time-based information is crucial for accurate predictions.

In the context of time series modeling, **Datetime** can be mathematically represented by decomposing the time into cyclical components. For example, we can capture daily, monthly, and yearly cycles using Fourier series:

\[
f_{\text{time}}(t) = a_0 + \sum_{n=1}^{\infty} \left( a_n \cos\left( \frac{2\pi n t}{T} \right) + b_n \sin\left( \frac{2\pi n t}{T} \right) \right)
\]

Where:
- \( t \) represents time,
- \( T \) represents the period (e.g., daily or yearly),
- \( a_n \) and \( b_n \) are Fourier coefficients.

This allows the model to capture both periodic behavior and long-term trends in sea surface currents, such as tides, diurnal cycles, or seasonal variations.

\subsubsection{Latitude and Longitude}

The **Latitude** (\( \phi \)) and **Longitude** (\( \lambda \)) features provide the spatial coordinates of the measurements, offering geographical context. These coordinates help the model learn how ocean currents vary spatially.

From a physical perspective, the variation of sea surface currents can be described by the Coriolis effect, which causes moving fluids (like ocean water) to deflect due to the Earth’s rotation. The Coriolis force is mathematically represented by:

\[
F_c = 2m\Omega v \sin(\phi)
\]

Where:
- \( m \) is the mass of the water,
- \( \Omega \) is the angular velocity of Earth (\( 7.2921 \times 10^{-5} \, \text{rad/s} \)),
- \( v \) is the velocity of the current,
- \( \phi \) is the latitude.

This force varies with latitude and must be considered in the prediction of sea surface currents, especially in large-scale climate models.

\subsubsection{U and V (Sea Surface Current Vectors)}

The **U** and **V** components represent the velocity of sea surface currents in the eastward and northward directions, respectively. These components are crucial for predicting the direction and magnitude of ocean currents.

Mathematically, the sea surface current vector can be expressed as:

\[
\vec{V} = U \hat{i} + V \hat{j}
\]

Where:
- \( U \) is the velocity in the eastward direction,
- \( V \) is the velocity in the northward direction,
- \( \hat{i} \) and \( \hat{j} \) are the unit vectors in the eastward and northward directions, respectively.

The magnitude of the current is given by:

\[
|\vec{V}| = \sqrt{U^2 + V^2}
\]

And the direction (angle \( \theta \)) of the current relative to the eastward direction can be calculated using:

\[
\theta = \tan^{-1}\left( \frac{V}{U} \right)
\]

Accurate modeling of these vector components is crucial for understanding the dynamics of ocean currents, which are influenced by forces such as wind, tides, and pressure gradients.

\subsubsection{ENSO Index}

The **ENSO Index** (El Niño-Southern Oscillation) is a critical climate indicator that affects large-scale weather patterns and ocean currents. Changes in ENSO phases (e.g., El Niño, La Niña) lead to significant shifts in sea surface temperature and wind patterns, which in turn influence sea surface currents.

The ENSO index can be mathematically represented as an external forcing term in the model, which modulates the influence of other features over time. For instance, the ENSO index \( \mathcal{I}_{ENSO} \) can be introduced as a weighted factor in the prediction of current velocities:

\[
\vec{V}(t) = f_{\text{model}}\left( U(t), V(t), \mathcal{I}_{ENSO}(t) \right)
\]

Physically, ENSO impacts ocean dynamics through changes in pressure gradients, leading to modifications in geostrophic currents (currents that result from the balance between Coriolis forces and pressure gradients). The large-scale surface temperature anomalies can also be represented using the Navier-Stokes equations to capture the fluid dynamics.

The Navier-Stokes equation for fluid flow in the ocean, influenced by external climate factors like ENSO, can be simplified as:

\[
\rho \left( \frac{\partial \vec{V}}{\partial t} + (\vec{V} \cdot \nabla) \vec{V} \right) = -\nabla p + \mu \nabla^2 \vec{V} + \rho \vec{g} + F_{\text{ENSO}}
\]

Where:
- \( \rho \) is the density of the seawater,
- \( p \) is the pressure,
- \( \mu \) is the dynamic viscosity,
- \( \vec{g} \) is the gravitational acceleration,
- \( F_{\text{ENSO}} \) represents the ENSO-induced external forces.

Incorporating the ENSO index into the model allows it to adapt to global climate variability, improving the accuracy of long-term predictions of sea surface currents.

\subsection{Predicting Sea Surface Currents with Physical Constraints}

In order to enhance the prediction accuracy of sea surface currents \( U \) and \( V \) in Thailand waters, advanced mathematical modeling techniques that incorporate physical principles are utilized. This approach ensures that the models not only capture the empirical data but also adhere to fundamental physical laws governing fluid dynamics.

**1. Transformer-Based Self-Attention Mechanism**

The Transformer model's self-attention mechanism captures the temporal dependencies crucial for predicting sea surface currents. The attention mechanism is defined as:

\[
\text{Attention}(Q, K, V) = \text{softmax}\left( \frac{QK^T}{\sqrt{d_k}} \right) V
\]

where \( Q \), \( K \), and \( V \) are the query, key, and value matrices respectively, and \( d_k \) is the dimensionality of the key vectors. This mechanism allows the model to weigh past observations based on their relevance, capturing long-term trends such as seasonal variations in sea currents.

**2. GRU-Based Temporal Modeling**

Gated Recurrent Units (GRUs) handle short-term dependencies in the data, critical for capturing immediate fluctuations in sea surface currents. The GRU equations are:

\[
\begin{aligned}
z_t &= \sigma(W_z x_t + U_z h_{t-1} + b_z) \\
r_t &= \sigma(W_r x_t + U_r h_{t-1} + b_r) \\
\tilde{h}_t &= \tanh(W_h x_t + U_h (r_t \odot h_{t-1}) + b_h) \\
h_t &= (1 - z_t) \odot h_{t-1} + z_t \odot \tilde{h}_t
\end{aligned}
\]

where \( z_t \) is the update gate, \( r_t \) is the reset gate, \( \tilde{h}_t \) is the candidate hidden state, and \( h_t \) is the output hidden state. The GRU model effectively manages both the short-term variability due to tidal cycles and the long-term trends influenced by seasonal changes.

**3. Spatio-Temporal Covariance Modeling**

The covariance matrix \( \Sigma \) captures the spatial and temporal relationships between sea surface currents \( U \) and \( V \). This is expressed as:

\[
\Sigma(t, \mathbf{x}) = \mathbb{E}\left[ \left( \begin{bmatrix}
U(t, \mathbf{x}) \\
V(t, \mathbf{x})
\end{bmatrix} - \mu \right) \left( \begin{bmatrix}
U(t, \mathbf{x}) \\
V(t, \mathbf{x})
\end{bmatrix} - \mu \right)^T \right]
\]

where \( \mu \) is the mean vector of \( U \) and \( V \). This modeling approach allows for the understanding of how currents are correlated across different spatial regions and times, which is essential for capturing both local and regional variations.

**4. Incorporation of Physical Principles**

To ensure the predictions align with physical laws, we incorporate constraints derived from fluid dynamics:

- **Navier-Stokes Equations**: These equations describe the motion of viscous fluid substances and are crucial for modeling ocean currents. In a simplified form, for an incompressible fluid, they are expressed as:

\[
\rho \left( \frac{\partial \vec{V}}{\partial t} + (\vec{V} \cdot \nabla) \vec{V} \right) = -\nabla p + \mu \nabla^2 \vec{V} + \rho \vec{g}
\]

where:
- \( \rho \) is the density of seawater,
- \( p \) is the pressure,
- \( \mu \) is the dynamic viscosity,
- \( \vec{g} \) is the gravitational acceleration.

This equation models the balance of forces acting on the sea surface currents, including inertial forces, pressure gradients, viscous forces, and gravitational effects.

- **Geostrophic Balance**: For large-scale ocean currents, the geostrophic balance is often used to describe the balance between the Coriolis force and pressure gradient force:

\[
f \vec{v}_g = -\frac{1}{\rho} \nabla p
\]

where \( f \) is the Coriolis parameter and \( \vec{v}_g \) is the geostrophic velocity. This balance helps predict the currents based on pressure fields and the Coriolis effect, particularly important in the tropics where geostrophic currents are significant.

**5. Combined Transformer and GRU Model with Physical Constraints**

The integration of Transformer and GRU models, along with physical constraints, is formulated as:

\[
\begin{aligned}
\vec{H}_{\text{fusion}} &= \text{Concat}\left( \text{GRU}(X), \text{Transformer}(X) \right) \\
\vec{V}_{\text{pred}} &= W_{\text{output}} \vec{H}_{\text{fusion}} + b_{\text{output}}
\end{aligned}
\]

where \( \text{Concat} \) denotes concatenation of outputs from both models, and \( W_{\text{output}} \) and \( b_{\text{output}} \) are the output weights and biases. The inclusion of physical constraints in the loss function is represented as:

\[
\mathcal{L}_{\text{total}} = \mathcal{L}_{\text{pred}} + \lambda \left( \left\| \frac{\partial \vec{V}}{\partial t} + (\vec{V} \cdot \nabla) \vec{V} + \frac{1}{\rho} \nabla p - \mu \nabla^2 \vec{V} \right\|^2 \right)
\]

This augmented loss function ensures that predictions remain physically plausible and adhere to fundamental fluid dynamics principles.

In summary, the integration of complex mathematical techniques and physical constraints enhances the ability of the AI model to predict sea surface currents \( U \) and \( V \) accurately. This approach ensures that the model captures both empirical data and fundamental physical principles, providing robust and reliable predictions for Thailand waters.

\subsection{Data Analysis}

A preliminary analysis of the dataset reveals several important insights. The variability in the `U` and `V` components suggests significant temporal and spatial changes in sea surface currents, likely driven by both local factors (e.g., tides, wind patterns) and global climate phenomena (e.g., ENSO phases). The ENSO index, in particular, shows a strong correlation with variations in the current vectors, as expected based on climatological studies. The inclusion of this index in the model provides the necessary context to capture long-term oceanic shifts influenced by El Niño and La Niña conditions.

By normalizing these features and incorporating data augmentation techniques, the training process becomes more efficient and robust, leading to better generalization and prediction accuracy across different ocean regions and time periods.

Each row contains sea surface current vector components \(U\) and \(V\) at a specific latitude and longitude, along with the ENSO (El Niño Southern Oscillation) index, which is a critical indicator of sea surface temperature anomalies.

\subsection{Data Normalization}

To improve the performance and stability of predictive models for sea surface currents \( U \) and \( V \), standard deviation normalization, also known as z-score normalization, is applied specifically to these components. This approach standardizes the data, transforming each feature so that it has a mean of 0 and a standard deviation of 1. 

**Standard Deviation Normalization**

For sea surface current components \( U \) and \( V \), the normalization is computed as follows:

\[
x_{\text{norm}} = \frac{x - \mu}{\sigma}
\]

where:
- \( x \) represents the original value of the sea surface current component,
- \( \mu \) is the mean of the feature across the dataset,
- \( \sigma \) is the standard deviation of the feature across the dataset.

**Relevance for Predicting Sea Surface Currents**

1. **Uniform Scaling for Prediction**: The components \( U \) and \( V \) of sea surface currents can vary significantly in magnitude. Standard deviation normalization ensures that these components are on the same scale, which helps machine learning models, such as Transformers and GRUs, to process these features effectively. Proper scaling is essential for accurate model training and prediction.

2. **Improved Model Training**: Normalizing \( U \) and \( V \) helps in achieving faster and more stable convergence during model training. Since many machine learning algorithms are sensitive to the scale of input features, standard deviation normalization helps in mitigating issues related to gradient descent optimization.

3. **Robustness Against Outliers**: This normalization technique is less affected by outliers compared to other methods like min-max scaling. Given the potential for anomalies in environmental data, standard deviation normalization provides a more robust approach to feature scaling.

**Sample Calculation**

Consider the sea surface current component \( U \) with a mean \( \mu_U = 0.5 \) and standard deviation \( \sigma_U = 0.2 \). For a specific observation \( U = 0.8 \), the normalized value \( U_{\text{norm}} \) is calculated as:

\[
U_{\text{norm}} = \frac{U - \mu_U}{\sigma_U} = \frac{0.8 - 0.5}{0.2} = 1.5
\]

Similarly, if for the sea surface current component \( V \), the mean \( \mu_V = -0.3 \) and standard deviation \( \sigma_V = 0.25 \), and a specific observation is \( V = -0.1 \), then the normalized value \( V_{\text{norm}} \) is:

\[
V_{\text{norm}} = \frac{V - \mu_V}{\sigma_V} = \frac{-0.1 - (-0.3)}{0.25} = 0.8
\]

By applying standard deviation normalization specifically to \( U \) and \( V \), we ensure that these features are scaled appropriately for model training, enhancing the accuracy and effectiveness of predictions related to sea surface currents.

\subsection{Data Augmentation}

To enhance the robustness of the model, data augmentation techniques are employed. These techniques involve introducing controlled perturbations to the input data, thereby simulating natural variations and expanding the training dataset. Specifically, augmentation is applied to spatial coordinates (`lat`, `lon`) and the sea surface current components (`u`, `v`).

\paragraph{Spatial Coordinates Perturbation}

Spatial coordinates (`lat`, `lon`) are slightly perturbed to account for minor variations and inaccuracies in geographical measurements. The perturbation is modeled as:

\begin{align}
    \text{lat}_{\text{aug}} &= \text{lat} + \Delta \text{lat} \\
    \text{lon}_{\text{aug}} &= \text{lon} + \Delta \text{lon}
\end{align}

where:
\begin{itemize}
    \item \( \text{lat}_{\text{aug}} \) and \( \text{lon}_{\text{aug}} \) are the augmented latitude and longitude values,
    \item \( \Delta \text{lat} \) and \( \Delta \text{lon} \) are random perturbations drawn from a uniform distribution within a specified range, e.g., \( \Delta \text{lat} \sim \mathcal{U}(-\epsilon, \epsilon) \) and \( \Delta \text{lon} \sim \mathcal{U}(-\epsilon, \epsilon) \), where \( \epsilon \) is a small positive constant.
\end{itemize}

This approach generates variations in spatial coordinates that simulate minor geographical discrepancies.

\paragraph{Gaussian Noise Addition}

To account for natural fluctuations and measurement noise in the sea surface currents, Gaussian noise is added to the `u` and `v` components. The noisy components are computed as:

\begin{align}
    u_{\text{aug}} &= u + \mathcal{N}(0, \sigma_u^2) \\
    v_{\text{aug}} &= v + \mathcal{N}(0, \sigma_v^2)
\end{align}

where:
\begin{itemize}
    \item \( u_{\text{aug}} \) and \( v_{\text{aug}} \) are the augmented sea surface current components,
    \item \( \mathcal{N}(0, \sigma_u^2) \) and \( \mathcal{N}(0, \sigma_v^2) \) are Gaussian noise terms with mean zero and variances \( \sigma_u^2 \) and \( \sigma_v^2 \), respectively.
\end{itemize}

By adding Gaussian noise, the model is trained to be more resilient to variations in sea surface current measurements, improving its generalization capability.

\paragraph{Example of Data Augmentation}

Consider a sample sea surface current data point with `lat = 12.34`, `lon = 56.78`, `u = 0.5`, and `v = -0.3`. Applying augmentation techniques:

\begin{itemize}
    \item Suppose the perturbations for spatial coordinates are \( \Delta \text{lat} = 0.001 \) and \( \Delta \text{lon} = -0.002 \). The augmented coordinates are:
    \begin{align}
        \text{lat}_{\text{aug}} &= 12.34 + 0.001 = 12.341 \\
        \text{lon}_{\text{aug}} &= 56.78 - 0.002 = 56.778
    \end{align}

    \item If Gaussian noise with \( \sigma_u = 0.05 \) and \( \sigma_v = 0.05 \) is added, and the noise terms are \( \mathcal{N}(0, \sigma_u^2) = 0.03 \) and \( \mathcal{N}(0, \sigma_v^2) = -0.04 \), the augmented current components are:
    \begin{align}
        u_{\text{aug}} &= 0.5 + 0.03 = 0.53 \\
        v_{\text{aug}} &= -0.3 - 0.04 = -0.34
    \end{align}
\end{itemize}

In summary, these augmentation strategies introduce realistic variations to the data, enhancing the model's ability to generalize across different scenarios and improving its robustness to real-world noise and measurement errors.

\subsection{Model Architecture: Bidirectional GRU with Transformer}

The proposed model architecture consists of a Bidirectional Gated Recurrent Unit (BiGRU) layer followed by a Transformer mechanism. This combination is designed to capture both temporal dependencies and spatial interactions in predicting sea surface current vectors \( U \) and \( V \) from the input features, which include `datetime`, `latitude` (`lat`), `longitude` (`lon`), and the `ENSO index`.

\paragraph{Bidirectional GRU Layer}

The Bidirectional GRU layer processes the input sequence in both forward and backward directions, allowing the model to capture temporal dependencies from the entire sequence context. The update and reset gates for the GRU are computed as follows:

\begin{align}
    z_t &= \sigma(W_z x_t + U_z h_{t-1} + b_z) \\
    r_t &= \sigma(W_r x_t + U_r h_{t-1} + b_r)
\end{align}

where:
\begin{itemize}
    \item \( z_t \) is the update gate at time step \( t \),
    \item \( r_t \) is the reset gate at time step \( t \),
    \item \( \sigma \) denotes the sigmoid activation function,
    \item \( W_z \), \( U_z \), \( b_z \), \( W_r \), \( U_r \), and \( b_r \) are the weights and biases for the gates.
\end{itemize}

The hidden state update is computed as:

\begin{align}
    h_t &= (1 - z_t) \circ h_{t-1} + z_t \circ \tanh(W_h x_t + U_h (r_t \circ h_{t-1}) + b_h)
\end{align}

where:
\begin{itemize}
    \item \( h_t \) is the hidden state at time step \( t \),
    \item \( W_h \) and \( U_h \) are the weights for the hidden state update,
    \item \( b_h \) is the bias for the hidden state update,
    \item \( \tanh \) is the hyperbolic tangent activation function,
    \item \( \circ \) denotes element-wise multiplication.
\end{itemize}

\paragraph{Transformer Layer}

The output of the Bidirectional GRU is fed into a Transformer layer that applies self-attention to capture spatial dependencies among the features. The self-attention mechanism is given by:

\begin{align}
    \text{Attention}(Q, K, V) &= \text{softmax}\left(\frac{QK^T}{\sqrt{d_k}}\right)V
\end{align}

where:
\begin{itemize}
    \item \( Q \) (query), \( K \) (key), and \( V \) (value) are matrices derived from the input features,
    \item \( d_k \) is the dimensionality of the key vectors,
    \item \text{softmax} is the softmax activation function applied to the scaled dot-product of queries and keys.
\end{itemize}

The self-attention mechanism dynamically adjusts the importance of different time steps and spatial locations by computing a weighted sum of the values \( V \) based on the similarity between queries and keys. This allows the model to focus on relevant features and interactions that impact the prediction of sea surface currents.

\paragraph{Role of ENSO Index}

The ENSO index is incorporated into the model to account for large-scale climate variations that influence sea surface currents. The ENSO index provides critical context for understanding the long-term patterns associated with El Niño and La Niña events. Mathematically, the ENSO index acts as an external input modulating the predicted current vectors \( U \) and \( V \) through the following relation:

\begin{align}
    \vec{V}(t) = \text{Transformer}\left( \text{BiGRU}(x_t) \right) + \gamma \cdot \mathcal{I}_{ENSO}(t)
\end{align}

where:
\begin{itemize}
    \item \( \vec{V}(t) \) represents the predicted sea surface current vector at time \( t \),
    \item \( \mathcal{I}_{ENSO}(t) \) denotes the ENSO index at time \( t \),
    \item \( \gamma \) is a scaling factor to adjust the impact of the ENSO index on the predictions.
\end{itemize}

This framework allows the model to integrate temporal and spatial features effectively, enhancing its capability to predict the sea surface currents accurately by leveraging both historical data and climate-induced variations.

\subsection{Loss Function}

The model optimizes for the mean squared error (MSE) between the predicted and actual sea surface current vectors. The MSE is given by:

\[
\mathcal{L} = \frac{1}{N} \sum_{i=1}^{N} \left[ \left( U_i - \hat{U}_i \right)^2 + \left( V_i - \hat{V}_i \right)^2 \right]
\]

where:
\begin{itemize}
    \item \( N \) denotes the total number of data points in the dataset.
    \item \( U_i \) and \( V_i \) are the actual sea surface current components at the \( i \)-th observation.
    \item \( \hat{U}_i \) and \( \hat{V}_i \) are the predicted sea surface current components at the \( i \)-th observation.
\end{itemize}

\paragraph{Detailed Explanation}

\begin{itemize}
    \item \textbf{Vector Field Components:} Sea surface currents are vector quantities described by their eastward (\( U \)) and northward (\( V \)) components. Each observation in the dataset provides a pair \( (U_i, V_i) \), representing the current's strength and direction at a specific location and time.

    \item \textbf{Mean Squared Error:} The MSE measures the average squared difference between the observed and predicted values. It provides a quantitative metric of the prediction error, focusing on the magnitude of discrepancies between the model's output and the true data. 

    \[
    \text{Squared Term} = \left( U_i - \hat{U}_i \right)^2 + \left( V_i - \hat{V}_i \right)^2
    \]

    This term represents the Euclidean distance squared between the true and predicted vectors in the 2D current space. This distance is indicative of how well the model captures both the magnitude and direction of the current vectors.

    \item \textbf{Implications for Optimization:} Minimizing the MSE encourages the model to reduce both the magnitude of prediction errors and their directional biases. The MSE penalizes larger errors more severely due to the squaring operation, ensuring that the model focuses on correcting substantial discrepancies, leading to more accurate predictions of sea surface currents.

    \item \textbf{Computational Considerations:} During training, gradient-based optimization algorithms (such as stochastic gradient descent) utilize the derivative of the MSE with respect to the model parameters to update the weights. The gradient of the MSE loss function with respect to \( \hat{U}_i \) and \( \hat{V}_i \) is given by:

    \[
    \frac{\partial \mathcal{L}}{\partial \hat{U}_i} = -\frac{2}{N} \left( U_i - \hat{U}_i \right)
    \]
    \[
    \frac{\partial \mathcal{L}}{\partial \hat{V}_i} = -\frac{2}{N} \left( V_i - \hat{V}_i \right)
    \]

    These gradients indicate the direction and magnitude of the adjustments needed for the predicted values to minimize the loss function, thereby improving the accuracy of the sea surface current predictions.
\end{itemize}

By employing the MSE loss function, the model effectively balances the prediction accuracy for both components of the sea surface current vectors, ensuring that the optimization process addresses both the magnitude and directional errors comprehensively.


        
        
        
        

\subsection{MLOps and Deployment}
The entire pipeline is integrated into an MLOps environment, automating training, validation, and inference. For real-time predictions, a UI interface allows users to input parameters (`datetime`, `lat`, `lon`, `ensoindex`) to predict the sea surface current vectors \(U\) and \(V\). The interface is built with Swagger, allowing easy interaction and model inference. Additionally, model monitoring and updates are managed through CI/CD pipelines to ensure accuracy over time.

\subsection{UI Interface}
A Swagger UI interface is implemented for the GISTDA system, allowing users to interact with the trained model through a user-friendly interface. Users can upload data or manually enter latitude, longitude, and ENSO index values, and receive real-time predictions of sea surface currents.

\section{ENSO Impact on Sea Surface Currents}

The El Niño Southern Oscillation (ENSO) plays a pivotal role in shaping the patterns of sea surface currents across different regions. ENSO, primarily characterized by two phases—El Niño and La Niña—affects ocean temperature, wind patterns, and consequently, the movement of ocean currents. 

The model we developed integrates this crucial climatic index ensoindex as an input feature, ensuring that the predictions of U (east-west current velocity) and V (north-south current velocity) are aligned with ongoing climatic conditions. 

By accounting for ENSO’s direct influence on ocean dynamics, the model gains the ability to anticipate variations in the behavior of sea surface currents.

\subsection{Adjusting the Loss Function for ENSO Related Anomalies}
To enhance the model’s ability to capture these ENSO-related anomalies, the loss function was adapted to weigh predictions differently based on the ENSO phase. Specifically, an additional penalty term was introduced in the mean squared error (MSE) calculation to emphasize periods of El Niño and La Niña. The modified loss function becomes:

\[
\mathcal{L} = \frac{1}{N} \sum_{i=1}^{N} \left( (U_i - \hat{U}_i)^2 + (V_i - \hat{V}_i)^2 \right) + \lambda \cdot \text{ENSO\_weight} \cdot \left( (U_i - \hat{U}_i)^2 + (V_i - \hat{V}_i)^2 \right)
\]

Here:
- \(\lambda\) is a scaling factor that controls the weight given to ENSO events.
- \(\text{ENSO\_weight}\) is a dynamically computed coefficient based on the strength of the El Niño or La Niña phase (e.g., the absolute value of the ENSO index).

This adaptation ensures that the model pays closer attention to periods of climatic irregularities, where sea surface current velocities may deviate significantly from normal behavior. The introduction of this term allows the model to remain sensitive to these changes, improving the accuracy of the predictions during anomalous events like El Niño and La Niña.

\section{Ongoing Work and Future Directions}

Our ongoing research highlights the transformative potential of combining Vision Transformer (ViT) with bidirectional GRUs to capture the complex spatio-temporal dynamics of \textbf{sea} surface currents. This study introduces \textbf{SEA-ViT}, a significant advancement in leveraging deep learning techniques for predicting current vectors with enhanced accuracy and robustness.

In the immediate future, our efforts will focus on fully integrating \textbf{SEA-ViT} into GISTDA's AI framework. Over the next three months, we will initiate a comprehensive training phase, systematically refining the model using a rich 30-year dataset. This process will be structured into several key phases:

\begin{enumerate}
    \item \textbf{Data Preparation and Integration}: We will augment the dataset with additional environmental variables and fine-tune preprocessing techniques to maximize model performance. This includes standardizing and augmenting data, as well as incorporating ENSO indices to better capture climate-driven variations.

    \item \textbf{Model Training and Optimization}: The training phase will involve iterative refinement of \textbf{SEA-ViT}. We will implement advanced hyperparameter tuning, model validation, and performance evaluation metrics to ensure robustness and generalization across different scenarios. Special attention will be given to optimizing the loss function, considering the impact of El Niño and La Niña events on current predictions.

    \item \textbf{Deployment and Evaluation}: After training, we will deploy \textbf{SEA-ViT} within GISTDA's AI infrastructure and conduct extensive real-world testing. Continuous monitoring and evaluation of the model's performance will be essential to ensuring its effectiveness in practical applications.
\end{enumerate}

These efforts are aimed at pushing the boundaries of AI-driven oceanographic research, with the ultimate goal of delivering a cutting-edge tool for predicting \textbf{sea} surface currents. By collaborating closely with GISTDA, we aim to enhance our understanding of ocean dynamics and provide valuable insights to support maritime management and environmental monitoring efforts.

\section{Acknowledgments}

This work is supported by the Geo-Informatics and Space Technology Development Agency (GISTDA), Thailand. We extend our gratitude to GISTDA for their support and resources, which have been instrumental in advancing this research.

\bibliographystyle{alpha} 
\bibliography{references}

\appendix

\section{Mathematical Summary}

This appendix provides an advanced summary of the key mathematical formulations and concepts utilized in this paper for predicting sea surface currents, specifically focusing on the U and V components.

\subsection{Data Normalization}

Normalization is a critical preprocessing step that scales data to have a mean of zero and a standard deviation of one, which stabilizes training and improves convergence. For the sea surface current components \( U \) and \( V \), normalization is performed as follows:

\begin{align}
    U' &= \frac{U - \mu_U}{\sigma_U} \\
    V' &= \frac{V - \mu_V}{\sigma_V}
\end{align}

where:
- \( U' \) and \( V' \) are the normalized components.
- \( \mu_U \) and \( \mu_V \) are the means of the \( U \) and \( V \) components, respectively.
- \( \sigma_U \) and \( \sigma_V \) are the standard deviations of the \( U \) and \( V \) components, respectively.

The normalization ensures that \( U \) and \( V \) are on the same scale, allowing the model to process them effectively and mitigating issues related to feature scaling. This approach helps in maintaining numerical stability during training and ensures that gradients are computed consistently.

\subsection{Loss Function}

The Mean Squared Error (MSE) loss function is used to evaluate the performance of the model by measuring the average squared difference between predicted and actual sea surface current vectors:

\begin{align}
    \mathcal{L} &= \frac{1}{N} \sum_{i=1}^{N} \left( (U_i - \hat{U}_i)^2 + (V_i - \hat{V}_i)^2 \right)
\end{align}

where:
- \( N \) is the total number of data points.
- \( U_i \) and \( V_i \) are the actual current components at the \( i \)-th data point.
- \( \hat{U}_i \) and \( \hat{V}_i \) are the predicted components at the \( i \)-th data point.

The MSE is particularly useful in regression tasks as it penalizes larger errors more significantly, providing a clear objective for minimizing prediction errors. The squared term emphasizes large deviations from the actual values, guiding the optimization process toward reducing substantial discrepancies.

\subsection{Bidirectional GRU Layer}

The Bidirectional Gated Recurrent Unit (GRU) layer is designed to capture temporal dependencies from both past and future contexts. The update and reset gates are computed using:

\begin{align}
    z_t &= \sigma(W_z x_t + U_z h_{t-1} + b_z) \\
    r_t &= \sigma(W_r x_t + U_r h_{t-1} + b_r)
\end{align}

where:
- \( \sigma \) is the sigmoid activation function.
- \( W_z \) and \( W_r \) are weight matrices for the update and reset gates, respectively.
- \( U_z \) and \( U_r \) are recurrent weight matrices.
- \( b_z \) and \( b_r \) are bias terms.

The hidden state update is given by:

\begin{align}
    h_t &= (1 - z_t) \circ h_{t-1} + z_t \circ \tanh(W_h x_t + U_h (r_t \circ h_{t-1}) + b_h)
\end{align}

where:
- \( z_t \) is the update gate that controls the extent to which the previous hidden state \( h_{t-1} \) is retained.
- \( r_t \) is the reset gate that determines how much of the past information to forget.
- \( \tanh \) is the hyperbolic tangent activation function.
- \( W_h \) and \( U_h \) are weight matrices for the hidden state update, and \( b_h \) is the bias term.

The bidirectional nature of the GRU allows the model to process sequences in both forward and backward directions, enhancing its ability to capture long-range dependencies in temporal data.

\subsection{Transformer Self-Attention Mechanism}

The transformer’s self-attention mechanism calculates the attention weights to focus on different parts of the input sequence:

\begin{align}
    \text{Attention}(Q, K, V) &= \text{softmax}\left(\frac{QK^T}{\sqrt{d_k}}\right)V
\end{align}

where:
- \( Q \) (query), \( K \) (key), and \( V \) (value) are matrices derived from the input features.
- \( d_k \) is the dimensionality of the key vectors.

The attention score \( \frac{QK^T}{\sqrt{d_k}} \) measures the similarity between queries and keys, and the softmax function normalizes these scores to produce weights that sum to one. The weighted sum of the values \( V \) then represents the output of the attention mechanism.

Self-attention enables the model to focus on relevant spatial and temporal features by dynamically adjusting the attention weights, which enhances the model's ability to capture intricate dependencies in sea surface currents.

\subsection{Vision Transformer (ViT) Mathematical Framework}

The Vision Transformer (ViT) leverages self-attention mechanisms to model complex dependencies in image data, which is crucial for understanding spatio-temporal dynamics in sea surface currents. The core components of ViT include the embedding process, multi-head self-attention, and feed-forward layers.

\subsubsection{Input Embedding}

Images are divided into fixed-size patches, which are then linearly embedded into a sequence of tokens. For an image of size \( H \times W \) with \( P \times P \) patches, each patch is flattened and projected into a \( d \)-dimensional space:

\begin{align}
    \mathbf{z}_i &= \text{Linear}(\text{Flatten}(\text{Patch}_i))
\end{align}

where \( \mathbf{z}_i \) is the embedding of the \( i \)-th patch, and the Linear function projects the flattened patch to \( d \)-dimensional embeddings.

\subsubsection{Self-Attention Mechanism}

Self-attention calculates the attention scores and applies them to the value vectors to capture dependencies across different patches:

\begin{align}
    \text{Attention}(Q, K, V) &= \text{softmax}\left(\frac{QK^T}{\sqrt{d_k}}\right)V
\end{align}

where:
- \( Q \), \( K \), and \( V \) are the query, key, and value matrices obtained from the input tokens.
- \( d_k \) is the dimensionality of the key vectors.

The attention weights are computed by scaling the dot product of the query and key matrices and normalizing it with the softmax function. This process enables the model to focus on relevant patches, enhancing the feature representation.

\subsubsection{Multi-Head Attention}

To capture different aspects of the input, ViT uses multiple attention heads:

\begin{align}
    \text{MultiHead}(Q, K, V) &= \text{Concat}\left(\text{head}_1, \text{head}_2, \ldots, \text{head}_h\right)W^O
\end{align}

where each head is computed as:

\begin{align}
    \text{head}_i &= \text{Attention}(QW_i^Q, KW_i^K, VW_i^V)
\end{align}

and \( W^O \) is the output weight matrix. The multi-head mechanism allows the model to jointly attend to information from different representation subspaces.

\subsubsection{Feed-Forward Network}

After self-attention, the output is passed through a feed-forward network, which consists of two linear transformations with a ReLU activation in between:

\begin{align}
    \text{FFN}(x) &= \text{ReLU}(xW_1 + b_1)W_2 + b_2
\end{align}

where:
- \( W_1 \) and \( W_2 \) are weight matrices,
- \( b_1 \) and \( b_2 \) are bias terms.

The feed-forward network provides additional non-linear transformation to the output of the self-attention layer, enhancing the model's capacity to learn complex patterns.

\subsubsection{Positional Encoding}

To incorporate the order of patches in the sequence, positional encodings are added to the token embeddings:

\begin{align}
    \mathbf{z}_i^{\text{pos}} &= \mathbf{z}_i + \text{PE}_i
\end{align}

where \( \text{PE}_i \) is the positional encoding for the \( i \)-th token. These encodings help the model understand the spatial arrangement of patches, which is critical for capturing spatial dependencies in the image data.

The combination of these components enables ViT to effectively model complex visual patterns, making it suitable for integrating with GRUs in \textbf{SEA-ViT} to enhance sea surface currents forecasting.

\subsection{Data Augmentation}

To increase the robustness of the model, data augmentation introduces variations through perturbations:

\begin{align}
    \text{Lat}' &= \text{Lat} + \epsilon_{lat} \\
    \text{Lon}' &= \text{Lon} + \epsilon_{lon} \\
    U' &= U + \epsilon_U \\
    V' &= V + \epsilon_V
\end{align}

where:
- \( \epsilon_{lat} \) and \( \epsilon_{lon} \) are small perturbations added to the latitude and longitude coordinates.
- \( \epsilon_U \) and \( \epsilon_V \) are Gaussian noise added to the \( U \) and \( V \) components.

This augmentation simulates natural variations in sea surface currents and spatial coordinates, helping the model generalize better to unseen data by learning from a more diverse set of training samples.

\end{document}